\documentstyle[seceq,epsf,twoside]{ptptex}
 
\newcommand{\bcen}{\begin{center}} 
\newcommand{\ecen}{\end{center}}

\newcommand{\sisi}{$\sigma\sigma$ } 
\newcommand{\pipi}{$\pi\pi$ } 
\newcommand{\kk}{$K\overline{K}$ }

\newcommand{\bc}{\begin{center}} 
\newcommand{\ec}{\end{center}}

\newcommand{\epsig}{$f_0(500)$ } 
\newcommand{\fo}{$f_0(980)$ } 
\newcommand{\epw}{$f_0(1400)$ }

\newcommand{\btablll}{\begin{tabular}{lll}} 
\newcommand{\btabllr}{\begin{tabular}{llr}} 
\newcommand{\btabll}{\begin{tabular}{ll}} 
\newcommand{\etab}{\end{tabular}} 
\notypesetlogo  
\title{Scalar meson spectroscopy: achievements and traps}
\author{Robert {\sc Kami\'nski}, Leonard {\sc Le\'sniak}$^{*}$ and 
Benoit {\sc Loiseau$^{**}$} }
\inst{%
$^{*}$Henryk Niewodnicza\'nski Institute of Nuclear Physics,\\ 
PL 31-342 Krak\'ow, Poland\\
$^{**}$LPNHE/LPTPE Universit\'e P. et M. Curie, 4, Place Jussieu,\\
75252 Paris CEDEX 05, France
}
\abst{%
Interactions in three coupled channels: \pipi, \kk and \sisi
have been investigated in a wide two-pion effective mass region 
from the \pipi threshold up to 1600 MeV.
Analytical structure of amplitudes in all channels has been studied.
It was shown that its knowledge is necessary to understand spectrum 
of scalar mesons and their nature.   
}

\begin{document}
\maketitle

\setcounter{tocdepth}{4}

\section{Introduction}

Scalar mesons play important role in description of low energy hadron physics.
Their spectrum and structure is, however, still not clear.
In last years the lightest scalar state \epsig (or $\sigma$) was a subject of intensive studies
\cite{pdg2000}. 
Another state $f_0(980)$ is often regarded as \kk or multi-quark bound state.   
Relatively well experimentally known $f_0(1500)$ is a candidate 
for~a~glueball. 

Interactions between pairs of light mesons are a main source of information about the
spectrum and the structure of scalar mesons.
In 
\cite{kll2} 
we have analysed interactions in three coupled channels: \pipi, \kk and \sisi
using a separable potential model in two-pion effective mass region from
the \pipi threshold up to 1600 MeV.
The set of 6 solutions was used to study the analytical structure of
amplitudes. 
Spectrum of scalar mesons and their properties (widths, cross sections, branching ratios, coupling constants)
were derived.    
  
\section{Model}

Solution of a system of Lippmann-Schwinger equations for the meson-meson interaction
amplitudes leads to evaluation of 
Jost's function $D(k_{\pi}, k_K, k_{\sigma})$ 
which depends on momenta in three coupled channels.
In a fully decoupled case (when all interchannel couplings are equal to zero)
the Jost function separates into a product of three independent Jost functions
\begin{equation}
D(k_{\pi}, k_K, k_{\sigma})= D(k_{\pi})D(k_K)D(k_{\sigma})
\label{jostepar}
\end{equation}
and for example the \pipi element of the $S$-matrix reads 
\begin{equation}
S_{\pi\pi} = \frac{D(-k_{\pi})}{D(k_{\pi})} = e^{2i\delta_{\pi}},
\label{spipisepar}
\end{equation}
where $\delta_{\pi}$ is the \pipi phase shift.
Due to general relation 
\begin{equation}
D(k_{\pi}, k_K, k_{\sigma}) = D^*(-k_{\pi}^*, -k_K^*, -k_{\sigma}^*)
\label{general}
\end{equation}
and relation (\ref{spipisepar})
zeroes of the numerator and the denominator 
($S_{\pi\pi}$ zeroes and poles, respectively) lie symmetrically 
with respect to the real axis in the complex momenta planes.
Hereafter they will be called original poles and zeroes. 
Cancellation of moduli of the numerator and the denominator leads to 
inelasticity 
$\eta = |S_{\pi\pi}| = 1$ what corresponds to elastic scattering.

In a coupled case (when some interchannel couplings are different from zero)
the diagonal $S$-matrix elements can be expressed as
ratios of two Jost functions. For example the $S_{\pi\pi}$ element
has a form:
\begin{equation}
S_{\pi\pi} = \frac{D(-k_{\pi}, k_K, k_{\sigma})}{D(k_{\pi}, k_K, k_{\sigma})}.
\label{spipifull}
\end{equation}
Presence of the interchannel couplings leads to 
a movement of the 
$S$-matrix zeroes and 
poles from their original positions to other positions in the coupled case.
Interchannel couplings lead also to splitting of an original $S$-matrix zero or pole 
into $2^{n-1}$ poles and zeroes ($n$ is a number of coupled channels).
The $S$-matrix singularities lie on sheets 
denoted by signs of imaginary parts of complex momenta
($Im k_{\pi} Im k_K Im k_{\sigma}$).
Some of those poles and zeroes in the coupled case can lie close enough to physical 
region to have a significant influence on phase shifts and inelasticities. 
Such singularities correspond to resonances with parameters related to
the pole positions $k_p$ by: 
\begin{equation}
4(Re k_p + i Im k_p)^2 + 4m_{\pi}^2 = M^2 - i M \Gamma,
\label{relation}
\end{equation}
where $M$ is mass of a resonance and $\Gamma$ is its width.

\section{Results}

\begin{figure}[h]
  \epsfxsize=14 cm
  \epsfysize=17 cm
 \centerline{\epsffile{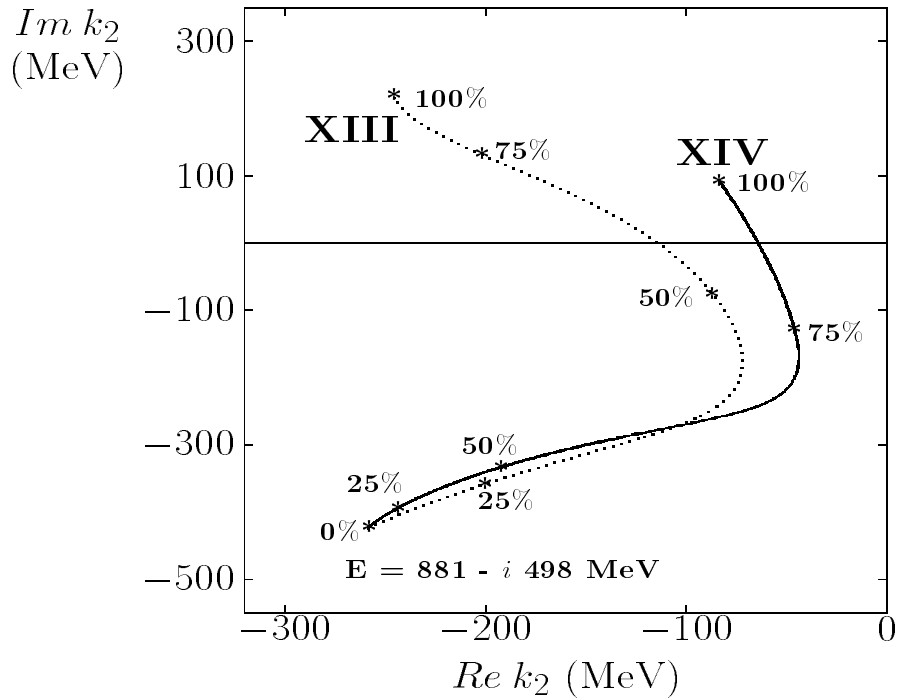}}
  
  \vspace{-4.8cm}
  
 \caption{Example of pole trajectories in the complex kaon momentum 
 plane as a function of the percentage of the interchannel coupling 
 strength for solution A in$^{2)}$.
  Roman numbers correspond to poles in Table 3 therein.}
  \label{fig:trace1}
\end{figure}

Knowledge of connections between positions of the $S$-matrix singularities in 
the fully coupled case and their original positions is necessary to investigate 
a spectrum and structure of scalar mesons. 
In our model we study positions of singularities as function of the interchannel couplings
which we gradually decrease starting from their values in the fully coupled case.
All values of the interchannel couplings are determined by our fits to the \pipi and \kk data
(see
\cite{kll2}).
In Fig. \ref{fig:trace1} one can see traces of two $S$-matrix poles for one of our solutions.
When all the interchannel couplings go to zero both poles change sheets on their ways
to their original position. 
One of them (number XIV) lies reasonably close to imaginary axis in the complex 
kaon momentum plane and may be treated as the \kk quasibound state. 
In Fig. \ref{fig:trace1} its original pole lies, however, 
below the real axis what is typical for an ordinary resonance.
It was pointed out in 
\cite{kll2} 
that a lack of precise experimental data near the \kk threshold leads to 
two possibilities for the \fo state: solutions with and without the \kk bound state
which have similar values of $\chi^2$ about 1 for one degree of freedom.

\begin{figure}[h]
  \epsfxsize=12 cm
  \epsfysize=16 cm
 \centerline{\epsffile{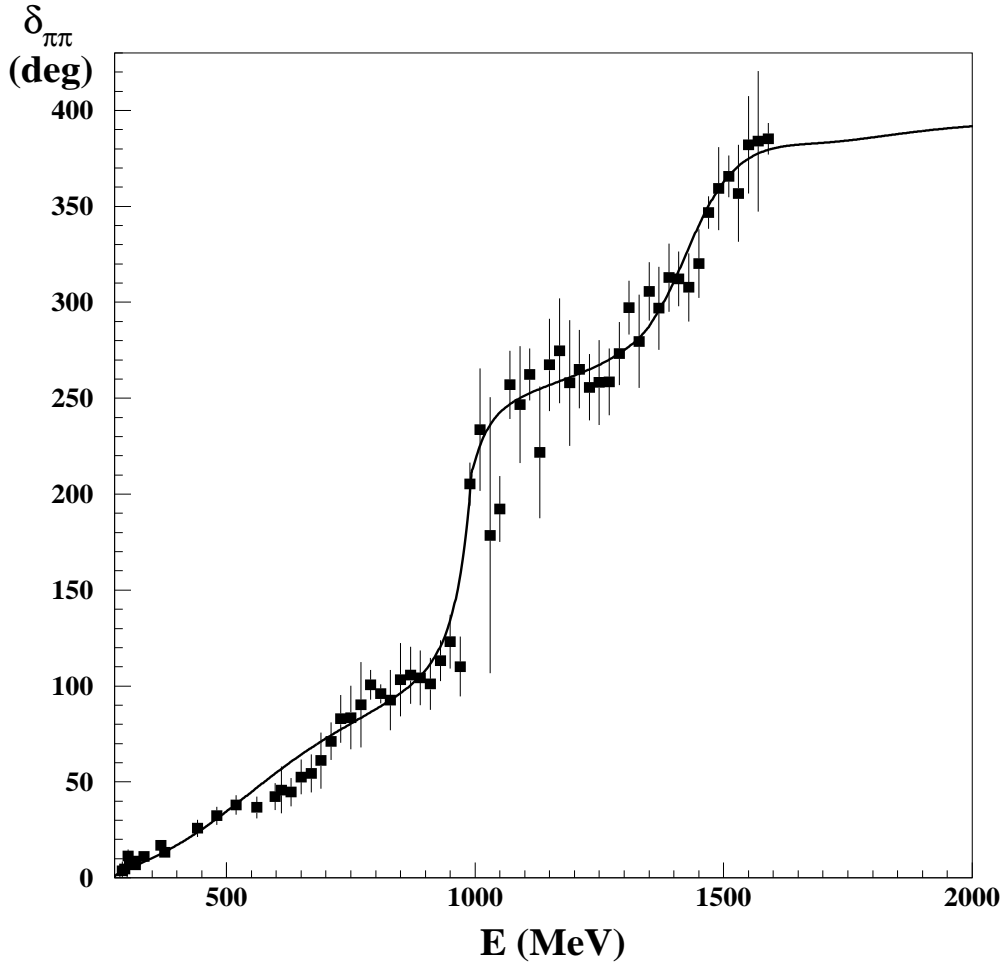}}
  
  \vspace{-2cm}
  
 \caption{Energy dependence of \pipi phase shifts for solution E}
  \label{fig:deltapi}
\end{figure}

In Fig. \ref{fig:deltapi} we show a comparison of data with another
solution up to 1600 MeV.
Above 1600 MeV one sees a flat behaviour of $\delta_{\pi\pi}$.
This can be explained by the fact that one $S$-matrix pole 
(number XIII in Table 5 in \cite{kll2})
 and the $S_{\pi\pi}$
zero (corresponding to pole number XVI there) lie on the same sheet 
at $m_{\pi\pi} = Re E \approx 1660$ MeV and reasonably near by the physical region.
Their close positions lead to an effective cancellation of their influence on the \pipi
phase shifts and 
explain why there is no apparent indication of any 
resonance in the theoretical curve in the region above 1600 MeV 
(see~Fig.~\ref{fig:deltapi}).

Finding of poles and zeros which have the most significant influence on 
phase shifts and inelasticities is important in identification of scalar 
resonances. 
All the $S$-matrix elements 
have the same denominator (see for example eq. (\ref{spipifull}) for $S_{\pi\pi}$)
so the $S$-matrix poles are common for all channels.
Zeroes of numerators have, however, different positions which depend
on a channel.    
In Table \ref{resonances} we present averaged masses and widths of
resonances found in our analyses (see
\cite{kll2}). 
They have been calculated for the $S$-matrix poles lying in proximity of the 
physical region.
Relatively small errors indicate a stability of the pole positions 
in all our solutions.
\begin{table}[h]
\centering
\caption{Average masses and widths of resonances $f_0(500)$, $f_0(980)$
and $f_0(1400)$ found in our solutions A, B, E and F from 
\cite{kll2}. Errors represent the maximum departure from the average.} 

\vspace{0.2cm}

\begin{tabular}{|c|c|c|c|}
\hline 
resonance & mass (MeV) & width (MeV) & sheet \\ 
\hline  
$f_0(500)$ or $\sigma$ & $523 \pm 12$ & $518 \pm 14$ & $-++$ \\ 
\hline 
$f_0(980)$ & $991 \pm 3$ & $71 \pm 14$ &  $-++$ \\ 
\hline 
& $1406 \pm 19$ & $160 \pm 12$ & $---$ \\ 
$f_0(1400)$ & $1447  \pm 27$ & $108 \pm 46$ & $--+$ \\ 
\hline  
\end{tabular} 
\label{resonances} 
\end{table} 
It can be seen in Table~\ref{resonances} that two important poles appear near 
the \sisi threshold. 
Although they lie on different sheets both should be taken into account
in determination of the \epw resonance parameters. 

\begin{figure}[h]
  \epsfxsize=15 cm
  \epsfysize=11.8 cm
 \centerline{\epsffile{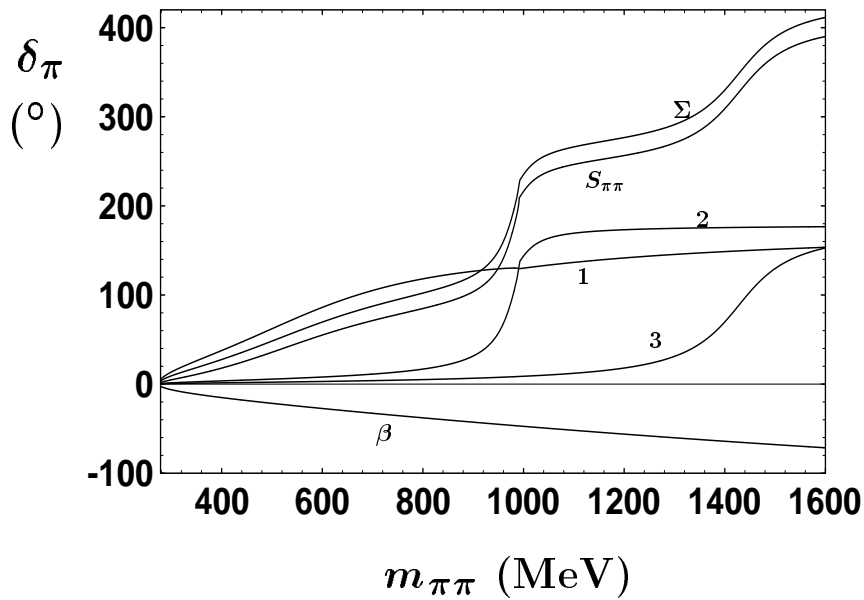}}
 
 \vspace{-1.8cm}
 
  \caption{Effective mass dependence of the \pipi phase shifts
  generated by particular poles and zeroes: 1 - related with $f_0(500)$,
 2 - related with $f_0(980)$ and 3 -  related with $f_0(1400)$, $\beta$ denotes
 part coming from the potential term, $\Sigma$ - sum of phases from all the 
 poles and $S_{\pi\pi}$ - full dependence of the \pipi phase shifts}
  \label{fig:sum}
\end{figure}

\begin{figure}[h!]
  \epsfxsize=7.5 cm
  \epsfysize=9.5 cm
  \centerline{\epsffile{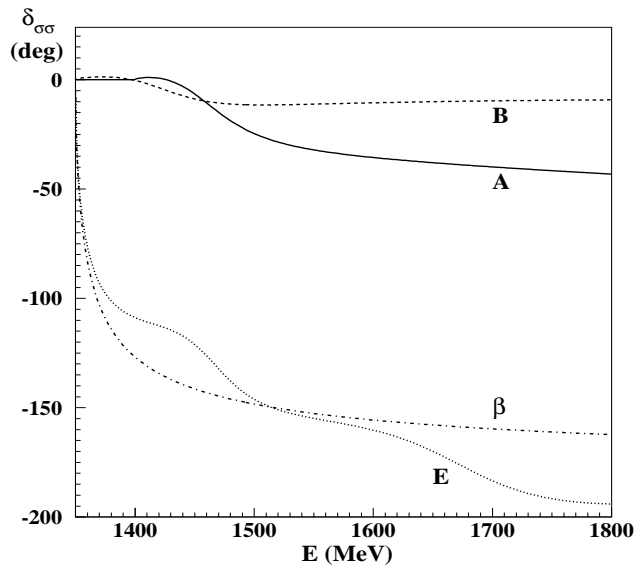}}
  
  \vspace{-1cm}
  
  \caption{Effective mass dependence of the \sisi phase shifts for solutions A, B and E.
  $\beta$ denotes the contribution of the potential term for solution E.}
  \label{fig:deltasigma}
\end{figure}

Knowledge of positions of all the $S$-matrix poles and zeroes allows us to describe 
behaviour of phase shifts 
(see analyses
\cite{kll2} and
\cite{klm}).
In Fig. \ref{fig:sum} one can see contributions corresponding to particular resonances
associated with single poles
and a part related to the double pole at $k = i\beta$ coming from the separable potential 
term of our  two-channel model~\cite{klm}.
The parameter $\beta$ is close to 1 GeV and the double pole contribution to 
$\delta_{\pi\pi}$ is decreasing monotonically down to about $-90^{o}$ at 
$k_{\pi} = \beta$ (at $m_{\pi\pi} \approx 2$ GeV). 
When the parameter $\beta$ is small the influence of the potential term on phase shifts 
in a particular channel is much stronger.
As it can be seen in Fig. \ref{fig:deltasigma} the double pole at $k_{\sigma} \approx
i 93$ MeV produces a very strong decrease of $\delta_{\sigma\sigma}$ phase shifts 
near the \sisi threshold for solution E in our three channel model
\cite{kll2} .
It is worthwhile to notice that a similar negative part called "background" has also 
been proposed by the Ishida group 
\cite{ishida}.

\section{Conclusions}

Analysis of analytical structure of the multichannel amplitudes is necessary to 
study scalar mesons.
Main achievement of that method is finding the $S$-matrix singularities which
have important influence on the phase shifts and inelasticities. 
The  poles can be related to physical resonances.
Near the thresholds more than one pole 
can play an important role in a given channel. 

Strong interchannel couplings can lead to significant movements of 
singularities and to a strong mutual cancellation of poles and zeroes lying 
reasonably close to physical region.
One of the main traps in analyses of the multichannel interactions 
is a difficulty to find unambiguous values of the resonance positions
by looking only at the phenomenological energy dependence of phase shifts and inelasticities.
Studies of $S$-matrix singularities are necessary to find parameters of resonances.

In our model total phase shifts can be composed of a sum of parts related to 
the amplitude poles and zeroes, cuts and the part coming from the potential term.
This last part is not arbitrarily added to the scattering amplitudes. 
It is created automatically 
by the double $S$-matrix pole and zero whose positions depend on the range parameters
of the separable potentials determined by fits to experimental data.

\acknowledgements

R. Kami\'nski would like to thank Prof. Shin Ishida and Dr. Muneyuki Ishida 
for an invitation and hospitality during the workshop.

\end{document}